\documentclass[runningheads]{llncs}

\usepackage{accv}

\usepackage{accvabbrv}

\usepackage{graphicx}
\usepackage{booktabs}

\usepackage[accsupp]{axessibility}  

\usepackage{hyperref}

\usepackage{orcidlink}

\usepackage{wrapfig}
\usepackage{multirow}

\begin{document}

\title{Parameter-Efficient Instance-Adaptive Neural Video Compression} 


\author{Seungjun Oh$^{\ast,1,\orcidlink{0009-0000-7989-6361}}$ \and
Hyunmo Yang$^{\ast,1,\orcidlink{0009-0008-3112-8572}}$\and
Eunbyung Park$^{\dagger,1,2,\orcidlink{0000-0003-4071-2814}}$}

\authorrunning{Oh and Yang et al.}

\institute{$^1$Department of Artificial Intelligence, Sungkyunkwan University
\\$^2$Department of Electrical and Computer Engineering, Sungkyunkwan University
}
\maketitle

\let\thefootnote\relax\footnotetext{$^{\ast}$ Equal contribution. Authorship order determined by coin flip.}

\let\thefootnote\relax\footnotetext{$^{\dagger}$ Corresponding author.}

\begin{abstract}
  Learning-based Neural Video Codecs (NVCs) have emerged as a compelling alternative to standard video codecs, demonstrating promising performance, and simple and easily maintainable pipelines. However, NVCs often fall short of compression performance and occasionally exhibit poor generalization capability due to inference-only compression scheme and their dependence on training data. The instance-adaptive video compression techniques have recently been suggested as a viable solution, fine-tuning the encoder or decoder networks for a particular test instance video. However, fine-tuning all the model parameters incurs high computational costs, increases the bitrates, and often leads to unstable training. In this work, we propose a parameter-efficient instance-adaptive video compression framework. Inspired by the remarkable success of parameter-efficient fine-tuning on large-scale neural network models, we propose to use a lightweight adapter module that can be easily attached to the pretrained NVCs and fine-tuned for test video sequences. The resulting algorithm significantly improves compression performance and reduces the encoding time compared to the existing instant-adaptive video compression algorithms. Furthermore, the suggested fine-tuning method enhances the robustness of the training process, allowing for the proposed method to be widely used in many practical settings. We conducted extensive experiments on various standard benchmark datasets, including UVG, MCL-JVC, and HEVC sequences, and the experimental results have shown a significant improvement in rate-distortion (RD) curves (up to 5 dB PSNR) and BD rates compared to the baselines NVC. Our code is available on \href{https://github.com/ohsngjun/PEVC}{https://github.com/ohsngjun/PEVC}.
\end{abstract}

\section{Introduction}
\label{sec:intro}

In the current digital landscape, we are experiencing unprecedented growth in video content consumption. Despite technological advancements providing us with high-speed internet and significant storage capabilities, efficient video compression technology still remains essential to the whole system. Video compression standards, including H.264~\cite{wiegand2003overview}, H.265~\cite{sullivan2012overview}, and H.266~\cite{bross2021overview}, have played a critical role in ensuring seamless multimedia experiences.

As an alternative approach to conventional standard codecs, data-driven learning-based video codecs have gained significant attention due to their promising compression performance. Numerous studies have adapted and redesigned deep neural networks to perform encoding and decoding tasks in place of manually crafted algorithms~\cite{lu2019dvc,li2021deep, agustsson2020scale, hu2022coarse}. By leveraging the power of neural networks to autonomously learn efficient representations of video signals and showing the potential of data-driven video codecs, they have sparked considerable interest in further research and development in this field. 

The limitations of Neural Video Codecs (NVCs) primarily stem from their reliance on training data. While a network trained on large-scale video datasets may exhibit good performance across a broad spectrum of video types, this assumption does not always hold true in real-world scenarios due to various reasons, such as the lack of diversity in training data, the presence of uncertainty in the optimization processes, and the limited expressibility of the neural networks, among others.
One effective method for enhancing the generalization performance is to further fine-tune the pretrained NVCs on the specific video instance. 
This approach, known as instance-adaptive fine-tuning, has improved compression performance across different neural codecs. Lu \etal ~\cite{lu2020content} suggested fine-tuning only the encoder since modifying the decoder parts requires quantization before sending the data to the receiver. This quantization significantly increases the transmitted bits, resulting in slower framerates at identical bitrates. Despite the drawback associated with decoder fine-tuning, Rozendaal \etal ~\cite{van2021instance} demonstrated that a comprehensive fine-tuning approach, encompassing both encoder and decoder, is often superior to encoder-only fine-tuning when only the difference between pretrained and fine-tuned decoder parameters are quantized and transmitted. This strategic approach, applicable to a broad spectrum of learned-based codecs, holds the potential to serve as a pivotal method for enhancing the overall performance of NVCs.

While instance-adaptive methods improve the performance of NVCs, it is noteworthy that achieving this enhancement requires updating additional model parameters and processing time to tailor the model to specific videos. However, fine-tuning the entire model parameters or networks imposes a substantial load on the process, resulting in longer encoding times and an increase in the amount of data bits to be transmitted. 
In this work, we propose a parameter-efficient instance-adaptive neural video compression method. More specifically, we suggest utilizing the widely recognized Low-Rank Adaptation (LoRA)~\cite{hu2021lora} method to efficiently update the pretrained neural networks. By freezing the parameters of pretrained networks and introducing a few trainable parameters, the proposed instance-adaptive method makes the fine-tuning process more efficient, significantly improving fine-tuning time compared to the previous full fine-tuning approaches. Moreover, as only a few parameters are updated during training, 
the amount of data required for transmission can be substantially reduced.
Furthermore, unlike the full fine-tuning instance adaptation method, the proposed LoRA-based instant-adaptive strategy exhibits a more stable and robust fine-tuning process.

Among the many well-established neural video codecs, we investigate the effectiveness of the proposed method on SSF~\cite{agustsson2020scale}. This particular method stands out for its high-quality compression performance and consists of a typical set of video compression modules, such as image compression, flow prediction, and residual compression. Consequently, the comprehensive research conducted in this work can be readily transferred to other neural video codecs. The following are the main contributions of this paper.

\begin{itemize}
  \item To the best of our knowledge, this work is the first effort to utilize the LoRA type of parameter-efficient fine-tuning method for video compression tasks.
  \item We have investigated LoRA variants, examining the methods and locations for integrating LoRA modules into well-established neural video codecs. 
  \item The extensive experimental results on many standard benchmark datasets show that the proposed approach has significantly improved the performance compared to the baseline methods.
\end{itemize}

\section{Related Work}
\label{sec:Related_Work}

\subsection{Neural Video Compression}

Building upon traditional coding methods like H.264 \cite{wiegand2003overview} and H.265 \cite{sullivan2012overview}, DVC \cite{lu2019dvc} introduces a novel architecture that incorporates optical flow for motion compensation and utilizes an encoder-decoder structure composed of convolution layers. This architecture compresses both residual information and motion derived from optical flow. Numerous subsequent studies have further enhanced this architecture with advanced techniques, including the use of multiple reference frames ~\cite{lin2020m, hu2021fvc}, recurrent auto-encoders and probability models~\cite{yang2020learning}, and contextual learning~\cite{li2021deep,sheng2022temporal,li2022hybrid,li2023neural, li2024neural}. As a well-known way for improving NVCs, motion compensation has been enhanced through the transition from optical flow to scale space flow \cite{agustsson2020scale}, deformable convolution ~\cite{hu2021fvc, yang2022learned}, or cross-scale flow~\cite{guo2021learning}.

Despite their advancements following traditional codec architecture, there are some studies where traditional codecs have been slightly or completely modified to further enhance performance. CANF-VC ~\cite{ho2022canf} and its subsequent studies ~\cite{chen2023b, hadizadeh2022lccm,chen2023canf} leverage augmented normalizing flow. MMVC ~\cite{liu2023mmvc} introduces different modes corresponding to the feature. Among various methods, we select online adaptation to overcome these challenges.

\subsection{Content-Adaptive Compression}
Neural data compression methods, trained on extensive datasets, can struggle with performance degradation when the data domain differs from the training set or if the data is exceedingly complex. To overcome this constraint, numerous studies fine-tuned the test data or out-of-domain data. Certain method optimize network without updating decoding parts~\cite{campos2019content, yang2020improving, djelouah2019content, abdoli2023gop, zhao2021universal, gao2022flexible, tang2024offline, xu2023bit}, which has shown promise for model optimization. Additional network application for domain transfer in Neural Image Compression~\cite{shen2023dec, tsubota2023universal, lv2023dynamic} has indicated that fine-tuning with test data domain can enhance compression quality. 

In video compression, Lu \etal ~\cite{lu2020content} demonstrated performance improvement by updating only the encoder network. Conversely, some research updates the decoding part. Rozendaal \etal ~\cite{van2021instance} employs entire model parameters for overfitting the test data and transmits the changes. This approach has resulted in an enhancement of the compression quality compared with their base models. However, the improvement was less significant in low-resolution videos because of the increased bit-rate cost. Research has also been conducted on updating only a portion of the parameters in the decoder network~\cite{zou2021adaptation, lam2020efficient}, offering a balanced approach between performance improvement and computational efficiency. 

Implicit Neural Representation (INR) methods have also attempted to optimize specific videos. NeRV~\cite{chen2021nerv} was one of the pioneers in integrating INR into the video compression pipeline. Due to its fast decoding time, NeRV was considered a potential replacement for traditional codecs. However, subsequent studies~\cite{li2022nerv, lee2023ffnerv, chen2023hnerv, zhao2023dnerv, he2023towards, kwan2024hinerv, xu2024vq}, despite aiming to enhance quality, have shown worse reconstruction performance compared to traditional methods.

\subsection{Parameter-Efficient Fine-Tuning }
As models continue to grow in size, the increasing computational costs and insufficient memory storage significantly hinder effective model training. The pioneer of adapters~\cite{rebuffi2017learning} introduces the injection concept to the architecture. Adapter~\cite{houlsby2019parameter}, comprised of down projection, up projection, and non-linear layers, is designed to construct a new branch sequentially between the pretrained layers. This sequential integration improves memory efficiency and reduces computation costs. Subsequent studies~\cite{chen2022vision, liu2022polyhistor, chen2022adaptformer} have expanded the application area of adapters, broadening the adapter architecture. These advances have yielded successful results in areas such as image compression~\cite{tsubota2023universal,shen2023dec}.

Despite these advantages, adapters encounter latency issues during inference, primarily due to the presence of non-linear operations. LoRA~\cite{hu2021lora} addresses this concern by eliminating the non-linear layer within the adapter module, which leads to better performance. Numerous LoRA-based researches ~\cite{he2023parameter,jie2023fact,luo2023towards, aleem2024convlora} have successfully extended their applications to various vision areas. Notably,~\cite{lv2023dynamic} establishes a connection between image compression and LoRA in the decoder. Our novel approach involves integrating the LoRA module into the CNN layer within the decoder of a video codec, a previously unexplored way.

\section{Method}
\subsection{Overview}
\begin{figure*}[t]
  \centering
   \includegraphics[width=\linewidth]{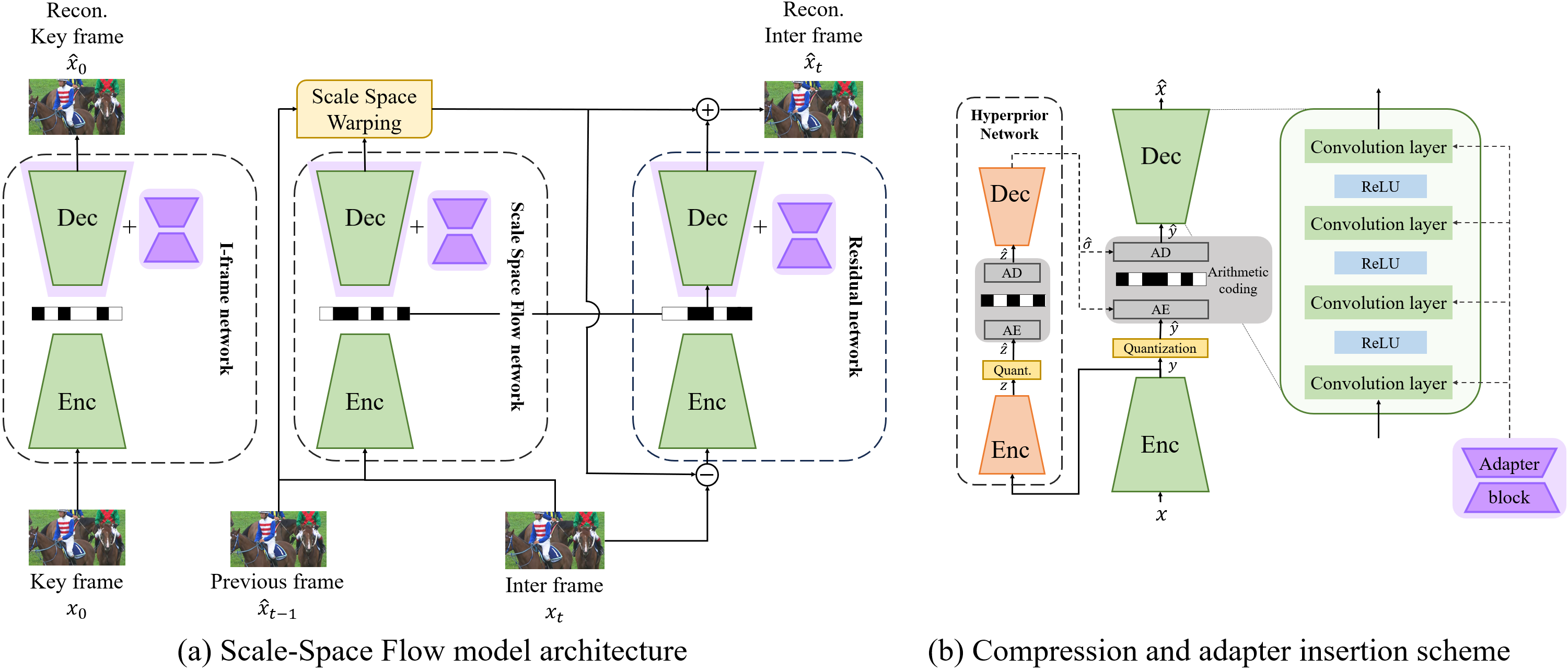}
   \caption{(a) provides an overview of SSF ~\cite{agustsson2020scale} decoding sequences. 
   (b) illustrates the structure of the compression model using hyperprior network ~\cite{balle2018variational}. AE and AD stand for Arithmetic Encoding and Arithmetic Decoding, respectively.}
   \label{fig:model}
\end{figure*}

In this section, we describe the proposed method, parameter-efficient instance-adaptive video compression. We employ the scale-space flow~\cite{agustsson2020scale} as our baseline model, which compresses the I-frames (Intra-coded frames, or key frames) and P-frames (Predicted frames) through neural networks. As shown in Fig.~\ref{fig:model}, the model consists of three main structures, each responsible for compressing different parts of the video: key frame, motion information, and residual information. Each encoder-decoder pair uses a hyperprior network to compress latent data. The proposed method inserts an adapter into the decoder layers, preserving the original model parameters while learning new ones to improve reconstruction for each video instance. The adapter parameters are then sent to the receiver, ensuring that the receiver can access a video of improved quality upon receipt.

\label{sec:Method}

\subsection{Preliminary: LoRA}

LoRA~\cite{hu2021lora} is a parameter-efficient fine-tuning technique for large neural networks. The core idea of LoRA is to train only a few model parameters during fine-tuning, making the fine-tuning more efficient without additional inference latency due to its linear operations. Given the frozen pretrained weight $W_0 \in \mathbb{R}^{C_{out} \times C_{in}}$, where $C_{out}$ and $C_{in}$ are the number of input and output channels, the trainable weight $\Delta W \in \mathbb{R}^{C_{out} \times C_{in}}$, which is used as an additional parallel branch layer, is introduced to find the optimal weight $W \in \mathbb{R}^{C_{out} \times C_{in}}$ by adjusting only a small subset of the parameters. LoRA assumes that the rank of $\Delta W$ is low, and as such, it is composed of a down-projection weight $A \in \mathbb{R}^{r \times C_{in}}$ and an up-projection weight $B \in \mathbb{R}^{C_{out} \times r}$, where $r \ll \texttt{min}(C_{in}, C_{out})$. Consequently, the weight matrix can be reparameterized as follows,
\begin{equation}
    W = W_0 + \Delta W = W_0 + AB.
\end{equation}

Only the low-rank matrices ($A$ and $B$) are trainable weights and the reparameterization incurs no additional latency during the testing inference.

\subsection{Adaptation Modules for Neural Video Codecs}
LoRA has been predominantly utilized in transformer architectures and attached to attention and linear layers. 
Since neural video codecs typically consist of multiple convolutional layers, we developed the revised LoRA architecture to make it compatible with these neural video codecs.

\subsubsection{LoRA in convolutional layers}
At first glance, incorporating the LoRA technique into convolutional layers may not seem challenging. Since convolution is a linear operator (it generally holds associativity and distributivity), we can define a LoRA module with two convolutional filters $A$ and $B$ along with the original convolutional filter $W_0$ as follows.
\begin{equation}
	W_0 \ast x + B \ast A \ast x = (W_0 + B \ast A) \ast x = (W_0 + \Delta W) \ast x,
\end{equation}
where $\ast$ is the convolution operator. Similar to how LoRA operates in fully connected layers, it can reduce the number of channels in the first layer (with the filter $A$), followed by the second convolutional layer (with the filter $B$) to have the same number of output channels as the original convolutional layer (with the filter $W_0$). Except for the non-linear activation functions in the middle, it shares similarities with the widely known bottleneck convolution block~\cite{resnet}.

However, the convolutions in deep neural networks easily break this assumption in practice due to various reasons, such as discrete convolution, small kernel sizes compared to the input features, and up (or down) samplings. For example, contemporary neural codecs have extensively used transposed convolutions to upsample the feature resolutions, and designing a LoRA module in such a scenario is not a straightforward task. 
In this work, therefore, we revised a rather simpler technique to efficiently represent the original convolutional kernels.

\subsubsection{Factorizing convolution kernels}
\begin{figure*}[!t]
\centering
    \begin{center}
    \includegraphics[width=\linewidth]{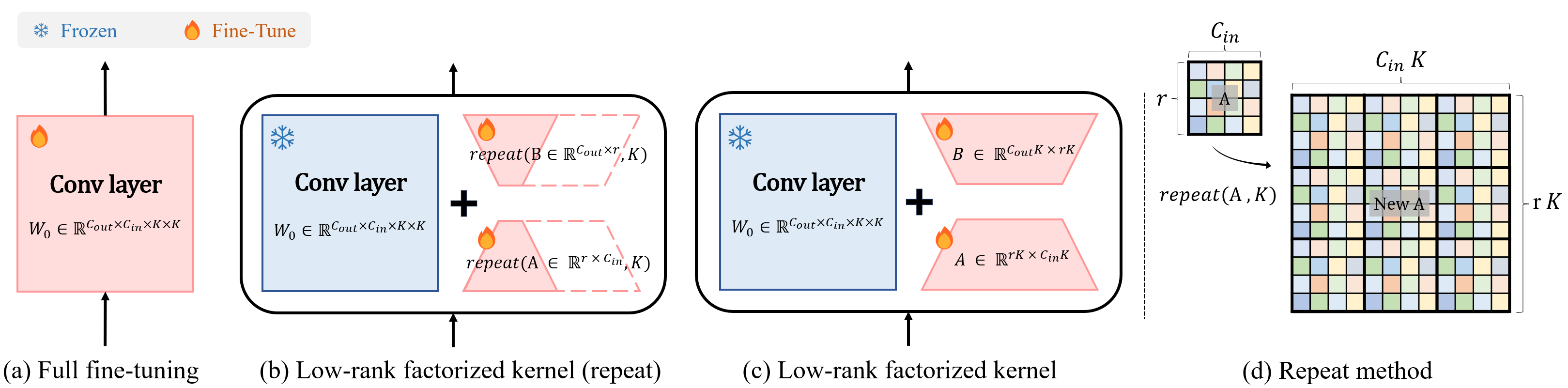}
    \end{center}
    \caption{Illustration of our proposed adapter architecture. (a) represents the original fine-tuning method that updates all parameters within the network. (b) and (c) only update the adapter network, where (c) uses more parameters than (b). (b) duplicates matrices according to the kernel size. (d) illustrates the repeat method applied in (b).}
    \label{fig:adapter} 
\end{figure*}

Let $W_0 \in \mathbb{R}^{C_{out} \times C_{in} \times K \times K}$ be a weight tensor for a convolutional layer, where $K$ represents the kernel size. Given the input feature $F_{in} \in \mathbb{R}^{C_{in} \times H \times W}$, where $H, W$ are height and width sizes, a convolutional layer linearly transforms $F_{in}$ into the output feature $F_{out} \in \mathbb{R}^{C_{out} \times H \times W}$. Similar to the original LoRA method, we introduce the trainable weight tensors $A \in \mathbb{R}^{r \times C_{in}}$ and $B \in \mathbb{R}^{C_{out} \times r}$, where $r \ll \texttt{min}(C_{in}, C_{out})$. Note that the number of training parameters is significantly smaller than the original parameters (e.g., $C_{in}, C_{out}=128, K=5, r=8$, we fine-tune only $0.5\%$ of parameters). To merge the original parameters and the newly introduced factorized matrices, we perform matrix multiplication and duplicate the resulting matrix to align its dimensions with those of the original parameters. More formally, the updated weight can be written as,
\begin{align}
    \widehat{A} = \texttt{repeat}(A,K),&\\
    \widehat{B} = \texttt{repeat}(B,K),&\\
    \texttt{repeat}(A, K) = A \otimes J_K,&\\
    W = W_0+\texttt{reshape}(\widehat{B}\widehat{A}),
\end{align}
where $\otimes$ denotes the kronecker product and $J_K$ represents $K \times K$ all-ones matrix. Hence, $\texttt{repeat}(\cdot, \cdot)$ copies the input matrix $K^2$ times and concatenates the duplicated matrices to construct an enlarged matrix as depicted in Fig.~\ref{fig:adapter}-(d), and the repeated matrices have the shapes of $\widehat{A} \in \mathbb{R}^{rK \times C_{in}K}$ and $\widehat{B} \in \mathbb{R}^{C_{out}K \times rK}$. Subsequently, we apply an operator  
$\texttt{reshape}(\cdot) : \mathbb{R}^{C_{out}K \times C_{in}K} \rightarrow \mathbb{R}^{C_{out} \times C_{in} \times K \times K}$
to construct a $\Delta W$ to be added to the original convolutional kernel $W_0$.

Despite its minimal parameter usage, the proposed factorization approach is remarkably efficient for compressing videos on a per-instance basis. However, the small number of update parameters sometimes results in limited performance improvements on some datasets and faces challenges in scenarios requiring high bitrate compression. Using a larger value for $r$ can easily increase the number of trainable parameters, but through empirical observation, we have noticed that merely raising the rank often does not improve the fine-tuning process.

\subsubsection{Extended size of factorized kernel}
The additional structure we propose largely mirrors the previous approach, with the key difference being that it does not duplicate the matrix but instead uses a larger number of parameters. With the slight abuse of notation, this method involves two matrices that decompose the weight of the convolution $W_0$ into $A \in \mathbb{R}^{rK \times C_{in}K}$ and $B \in \mathbb{R}^{C_{out}K \times rK}$, where $K$ continues to represent the kernel size, consistent with the previous structure. The remaining elements of the structure are configured similarly to the previous setup. The updated parameter $W$ can be written as,
\begin{equation}
    W = W_0 + \texttt{reshape}(BA).
\end{equation}
The proposed method starts to adjust the adapter parameters for a video instance and freeze the pretrained parameters. By initializing the adapter parameters to zero, the model replicates the output of the original model at the beginning of the fine-tuning process.

\subsection{Instance-Adaptive Fine-Tuning}

Given a video instance during testing time, the proposed method performs fine-tuning by updating the proposed factorized convolutional kernels. After a few training iterations, the updated weights are quantized and compressed before transmission to the receiver. To maximize the compression ratio, they are also entropy-coded to the bitstream with a prior, along with the latent codes generated by the encoder. On the receiver side, it already has the pretrained decoder parameters and updates the convolutional kernels. The resulting model architecture remains identical to the pretrained model, hence no additional latency during decoding. 

\subsubsection{Decoder-only updates}
\label{subsedc:Deconly}
Modern NVCs have utilized the encoder and decoder architectures, where the encoder extracts the codes from the input videos, and the spatial resolution is downsampled along the feature extraction process. On the other hand, the decoder upsamples the extracted codes to reconstruct the videos up to the original resolution. While it is possible to fine-tune both the encoder and decoder, our empirical observations indicate that competitive performance can be achieved by fine-tuning only the decoder. Fine-tuning the encoder results in a slight improvement in the compression ratio, but it requires a longer training time due to the need for backpropagation operations down to the input layer. Similar results have been demonstrated by Rozendaal \etal ~\cite{van2021instance}, showing the limitations of encoder updates in the case of full fine-tuning. We provide experimental results in the~\cref{sec:ablation study}.

\subsubsection{Vanishing parameters}
\label{subsec:vanishing}
In the fine-tuning stage, we observed that a significant amount of update parameters are vanishing in the quantization process. This results in substantial performance degradation on the receiver side. The primary cause of this issue was the slight modifications made to the parameters, which can be lost during quantization by falling into the zero bin. To prevent significant information loss, we use higher learning rates to encourage larger updates to the parameters. This enhances compression performance and significantly reduces training time since we achieve good reconstruction quality in fewer training iterations.

\subsubsection{Rate-distortion fine-tuning}
The training loss optimized during the training is given by the following equation:  \begin{equation}
    L = \frac{1}{N} \sum_{i=0}^{N-1}{\lambda D(x_i, \hat{x}_i) + H(z_i)} + \beta H(\bar{w}),
\end{equation}
where $\lambda$ is trade off between $D$ and $H$, $D$ denotes distortion loss between ground truth $x$ and reconstructed data $\hat{x}$, and $H$ is entropy estimation to represent the compressed latent information from I-frame, motion, and residual. As $D$ can be either mean square error (MSE) or structural similarity index measure (SSIM), we use $D$ as MSE loss. $H(\bar{w})$ with coefficient $\beta$ represents the number of bits of the quantized factorized kernel weights. Following Rozendaal \etal ~\cite{van2021instance}, we set spike-and-slab prior~\cite{johnstone2009statistical} to estimate the entropy estimation of the factorized kernel weights.
\begin{equation}
    p(w) = \frac{\mathcal{N}(w|0, \sigma^{2} I)+\alpha \mathcal{N}(w|0, s^2 I)}{1+\alpha},
\end{equation}

where $\sigma^2$ and $s^2$ denotes the variances of slab and spike components, respectively, and $\alpha$ is tunable parameter to set the scale of spike prior. The slap component keeps a lower scale of updates, while spike makes zero-update, enabling cheaper and sparser updates.

\section{Experiments}
\label{sec:Experiments}

\subsection{Experimental Setup}
\subsubsection{Dataset} We evaluate the performance of our method on the UVG-1k dataset \cite{mercat2020uvg}, MCL-JCV dataset~\cite{wang2016mcl}, and the HEVC class B and C dataset~\cite{bossen2013common}. We use RGB format for all video sequences. The resolution of the UVG dataset, MCL-JCV dataset, and the HEVC class B dataset are $1920 \times  1080$, with 7, 30, and 5 videos in each dataset, respectively. Due to our backbone model, which requires input sizes with a height and width that are multiples of 128, we crop the video resolution to $1920 \times 1024$. The HEVC class C dataset consists of 4 videos with a resolution of $832 \times 480$. For this dataset, We pad the right and bottom sides of the input image 
to fit into the model and crop outputs to obtain the final video.

\subsubsection{Video adaptation} We train each video using an MSE-optimized pretrained SSF model~\cite{agustsson2020scale}. We use nine qualities, each trained along to bitrate. We set $\lambda$ as $0.01 \cdot 2^{i}$, where $i$ ranges from -3 to 5, following the original approach. Both full fine-tuning and adapter fine-tuning methods are applied to train our model. We set $\alpha$ to 1000 and $\beta$ to 1 for loss calculation. The baseline model has 4 layers on the decoder, and we set the rank of the adapter $r$ as 16, 8, 8, and 2, starting from the first layer. Regardless of the frame length of the video data, we conducted 15 epochs of training for all video sequences. For the learning rate, we use 0.0001 for full fine-tuning and 0.0005 for our approach. We observed that using a high learning rate for full fine-tuning led to poorer results as training progressed, so we selected a value that could stably improve performance. Additionally, we use the  `ReduceOnPlateau'  learning rate scheduler. Since we use a higher learning rate than that used in training and train on a single video, once the model quickly converges, we reduce the learning rate to allow further training progress.
We set the Group of Pictures (GoP) to 4 during training, with batch size 3. For testing, we set the GoP to 12. Previous study~\cite{agustsson2020scale} have reported that training with smaller GoPs can lead to quicker convergence, and this strategy was also applied to instance-adaptation. Using smaller GoPs, rather than learning the testing sequences as they are, was beneficial for rapid convergence.

\begin{figure*}[!t]
\centering
    \begin{center}
    \includegraphics[width=\linewidth]{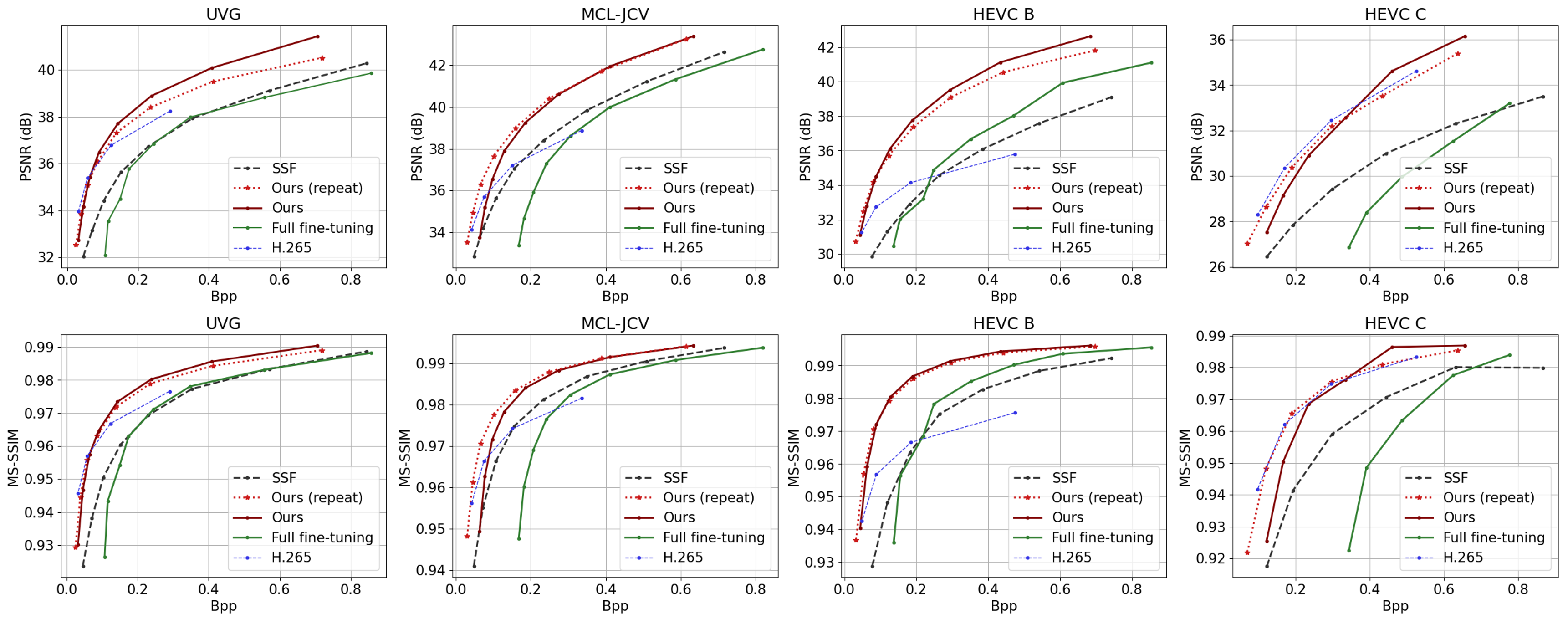}
    \end{center}
    \caption{Rate-Distortion curve comparison with the baseline method, SSF~\cite{agustsson2020scale} on UVG, MCL-JCV, HEVC class B, and C datasets}
    \label{fig:rdcurve} 
\end{figure*}

\begin{table}[t]
\centering
\setlength{\tabcolsep}{10pt}
\resizebox{\columnwidth}{!}{%
{\scriptsize
\begin{tabular}{@{}c|ccccc@{}}
\toprule
Method          & UVG    & MCL-JCV & HEVC B & HEVC C & Avg. \\ \midrule
SSF        & 111.26 & 22.40   & 50.63  & 121.10 & 76.35   \\
Full fine-tuning & 137.03 & 101.11  & 52.99  & 167.35 & 114.62  \\
Ours (repeat)          & -0.14  & \textbf{-35.85}  & \textbf{-50.42} & \textbf{9.26}   & \textbf{-19.29}  \\
Ours            & \textbf{-6.48}  & -11.14  & -47.47 & 14.78  & -12.58  \\ 
\bottomrule
\end{tabular}%
}
}
\caption{BD-rate (\%) comparison with x265. A lower value indicates better performance compared to the reference codec.}
\label{tab:bdrate}
\end{table}

\subsection{Experimental Result }

\subsubsection{Quantitative Results} We evaluate the performance with Rate-distortion (RD) curve and BD-rate, anchoring on x265. The data for this codec was obtained from publicly available data online~\cite{Hu2020pvc}.
\cref{fig:rdcurve} shows the RD curve measured for RGB PSNR and MS-SSIM on the UVG, MCL-JCV, and HEVC class B and C datasets. Both methods we proposed significantly improved the performance of the existing model. While the instance adaptive method shows relatively weak performance on the low resolution data~\cite{van2021instance}, our proposed methods also demonstrate similar performance improvement on the HEVC class C dataset, which has smaller resolution compared to other datasets.

\cref{tab:bdrate} presents the BD-rate results. We also conducted tests on the UVG, MCL-JCV, and HEVC class B and C datasets. Our method shows significant improvement compared to the original result, and the result on the 1K dataset outperformed x265. Remarkably, the BD-rate of ours with duplication is the best due to significantly fewer trainable parameters, resulting in a lower size of sending bits.

\subsubsection{Qualitative Results} 
\begin{figure*}[t]
  \centering
   \includegraphics[width=\linewidth]{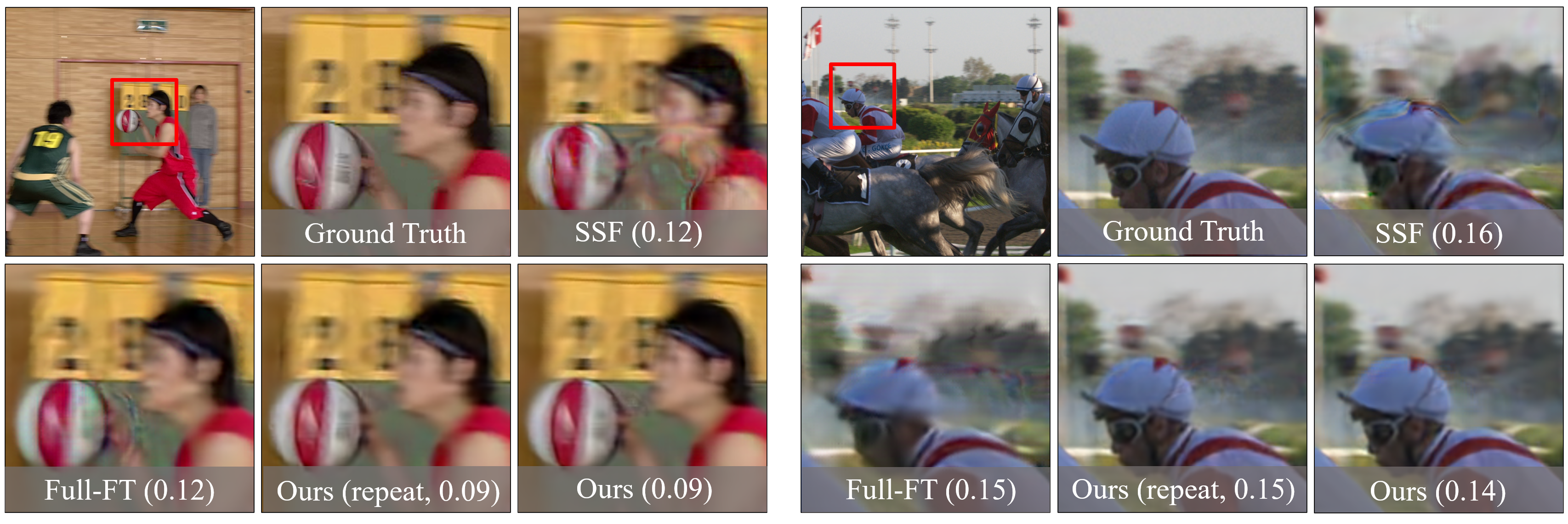}

   \caption{Qualitative result for moving object, numbers in the pictures represent the corresponding bpp. Our method effectively reduces distortions in P frames, resulting in a clearer and more accurate representation in similar bpp. (Left) 'BasketballDrive' sequence from the HEVC class B dataset. (Right) 'ReadySteadyGo' sequence from the UVG dataset.}
   \label{fig:qualitative}
\end{figure*}

\cref{fig:qualitative} presents the video compression results. We compared the reconstructed frames with similar bpp. The fine-tuning method has shown effective adaptability to each video sequence. It was observed that by adopting the instance-adaptive method, we could achieve outputs that closely resemble the original. When compared to the original, significant improvements were observed with reduced motion blur, color distortion, and other artifacts in moving objects. However, performance improvements were not noticeable in full fine-tuning. On the left of the \cref{fig:qualitative}, it is evident that the SSF method distorts color and retains motion information. In contrast, these errors are mitigated in the instance-adaptive method. Notably, our method reduces degradations, particularly on the ball and human face.

\subsection{Ablation Studies}
\label{sec:ablation study}

\subsubsection{Encoder adaptation} Encoder-only updates exhibit limited improvement, as discussed in \cref{subsedc:Deconly} and Rozendaal \etal ~\cite{van2021instance}. We performed an experiment involving attachment to both the encoder and decoder, similar to full fine-tuning but with a smaller number of updating parameters. As shown in \cref{tab:my-table}, our method using both encoder and decoder without duplication shows lower BD-rates and faster training time compared to full fine-tuning, similar with using only decoder side. However, attaching both sides has a slight gap and longer training time compared to decoder-only training, leading us to opt for training only the decoder part.

\begin{table*}[t]
\centering
\setlength{\tabcolsep}{3pt}
\renewcommand{\arraystretch}{1.3}
\resizebox{\linewidth}{!}{%
{\large
\begin{tabular}{c|ccccccc|c}
\hline
\multirow{2}{*}{Methods} & \multicolumn{7}{c|}{BD-rate (\%)}  & \multirow{2}{*}{Training time (min)} \\ 
\cline{2-8}
                  & Beauty\; & ReadySteadyGo\; & Bosphorus\; & HoneyBee\; & Jockey\; & ShakeNDry\; & YachtRide &                                      \\ \hline
                  \midrule
Full fine-tuning   & 91.21  & -13.89  & 39.98 & 126.56 & 41.05 & 59.36 & 15.50     & 23                                   \\ \hline
Ours (repeat)            & \multicolumn{1}{c}{-13.89} & \multicolumn{1}{c}{-62.99}        & \multicolumn{1}{c}{-64.50}    & \multicolumn{1}{c}{-28.68}   & \multicolumn{1}{c}{-41.19} & \multicolumn{1}{c}{-14.95}    & -52.85    & 14                                   \\ \hline
Ours              & \multicolumn{1}{c}{-28.53} & \multicolumn{1}{c}{-67.52}        & \multicolumn{1}{c}{-64.33}    & \multicolumn{1}{c}{-47.66}   & \multicolumn{1}{c}{-45.84} & \multicolumn{1}{c}{-16.33}    & -56.92    & 14                                   \\ \hline
Ours (enc, dec)     & \multicolumn{1}{c}{-34.44} & \multicolumn{1}{c}{-66.74}        & \multicolumn{1}{c}{-67.71}    & \multicolumn{1}{c}{-53.17}   & \multicolumn{1}{c}{-50.38} & \multicolumn{1}{c}{-21.77}    & -59.00    & 16                                   \\ \hline

\end{tabular}%
}
}
\caption{BD-rate (\%) of each data in UVG dataset (with SSF as the anchor), and training time (in Minutes) of one video of $1920 \times 1024$ resolution and 600 frames. }
\label{tab:my-table}
\end{table*}

\subsubsection{Instance-adaptive environments} 
\begin{figure*}[t]
  \centering
   \includegraphics[width=\linewidth]{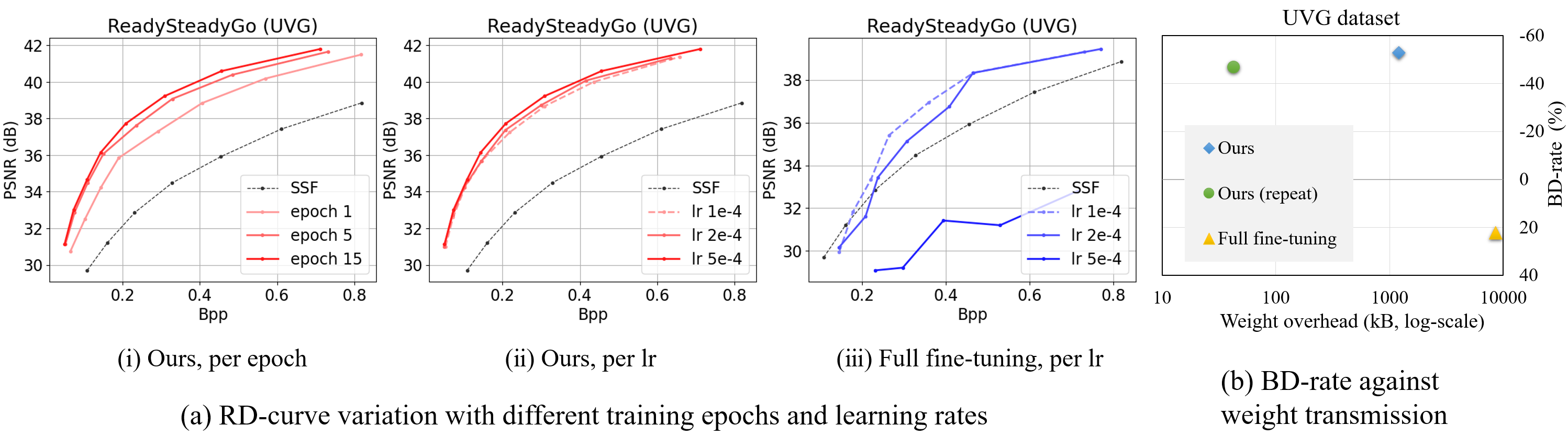}

   \caption{(a): RD curve varies according to training parameters, as measured on the ReadySteadyGo Dataset. (b): BD-rate (with the anchor as SSF) per adapter weight needed for decoding a video. In (b), while the BD-rate of Ours(repeat) and Ours is quite similar, Ours(repeat) requires significantly less overhead at only 43kB, compared to Ours which requires 1025kB, and full fine-tuning which necessitates 8522kB.}
   \label{fig:rd_lora_epoch}
\end{figure*}

We conducted an ablation study on various learning rates using ReadySteadyGo, as depicted in \cref{fig:rd_lora_epoch}. (\lowercase\expandafter{\romannumeral1}) We assessed the convergence speed in our predefined setting. After just one epoch, the curve exhibits a notable difference from the baseline, and the interval progressively narrows as epochs progress, signifies a fast convergence speed. (\lowercase\expandafter{\romannumeral2}) As mentioned in \cref{subsec:vanishing} concerning gradient vanishing, we trained our method with a learning rate ranging from 0.0001 to 0.0005. A slight expansion in the RD curve is observed, considered as quantization loss, which impedes faster convergence. (\lowercase\expandafter{\romannumeral2}, \lowercase\expandafter{\romannumeral3}) The results of our proposed method and full fine-tuning across various learning rates are presented. RD curves of full fine-tuning notably decline as the learning rate is increased. This suggests that full fine-tuning requires training at a lower learning rate, taking a longer time for convergence. On the other hand, using only factorized kernel during fine-tuning maintains performance regardless of the learning rate, and achieves this in a shorter time. Therefore, our proposed method demonstrates robustness across various hyperparameter environments.

\subsubsection{Additional decoder bits}
Before using the video codec, it is necessary to update fine-tuned information to synchronize the transmitter and receiver side. After this overhead transmission step is completed, video compression can be performed as in a typical NVC. 
\cref{fig:rd_lora_epoch} (b) demonstrates the quantized decoder overhead with BD-rate. Full fine-tuning requires transmitting 10 times or more weights compared to our proposed methods, despite the increased BD-rate. The repeated method, with significantly fewer trainable parameters, exhibits lower weight overheads and a reduced size of total bits. This is due to its reduced weight to be transmitted, resulting in a decreased total number of bitstreams. On the other hand, the non-repeated method shows the highest BD-rate among the three methods, even though the transmitted weight ratio relative to the overall bitstreams is larger than the repeated method. These results imply that our methods require a smaller weight bit size for transmission while achieving superior video compression.

\begin{wrapfigure}{r}{0.45\textwidth}
\setlength{\belowcaptionskip}{-30pt}
  \begin{center}
   \includegraphics[width=\linewidth, trim=0 0 0 200]{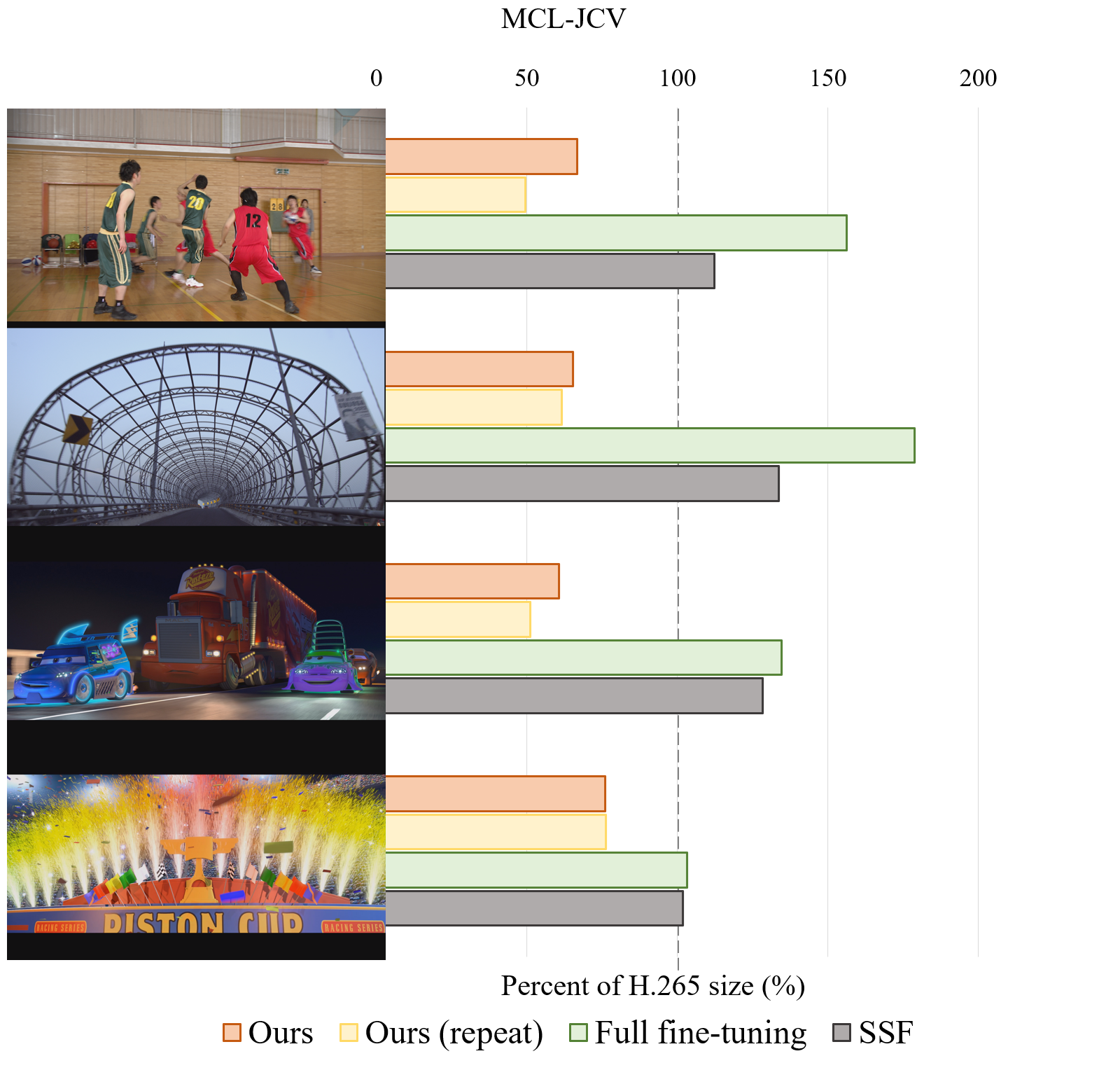}

   \caption{Rate savings for out-of-domain data. Y axis represents the bitrate compared to H.265(smaller is better).}
   \label{fig:domain}
   \end{center}
\end{wrapfigure}

\subsubsection{Hard-to-compress data} 
\label{subsec:hardcompress}
Agustsson \etal ~\cite{agustsson2020scale} acknowledged a failure to produce reasonable results on animation style datasets, which are out-of-domain. ~\cref{fig:domain} illustrates the size of encoded data relative to H.265, indicating the proportion of data used to compress a video with an equivalent PSNR. We selected two complex datasets and two cartoon style 
datasets which yield worse result compared to the traditional codec. Our proposed method has demonstrated its effectiveness in reducing the bitrate required for compression. Therefore, our methods enhance the motion estimation performance and exhibit better generalization across various video domains.

\subsubsection{Apply to another NVC} 
\begin{figure}[t]
\centering
   \begin{center}
       \includegraphics[width=1.0\linewidth]{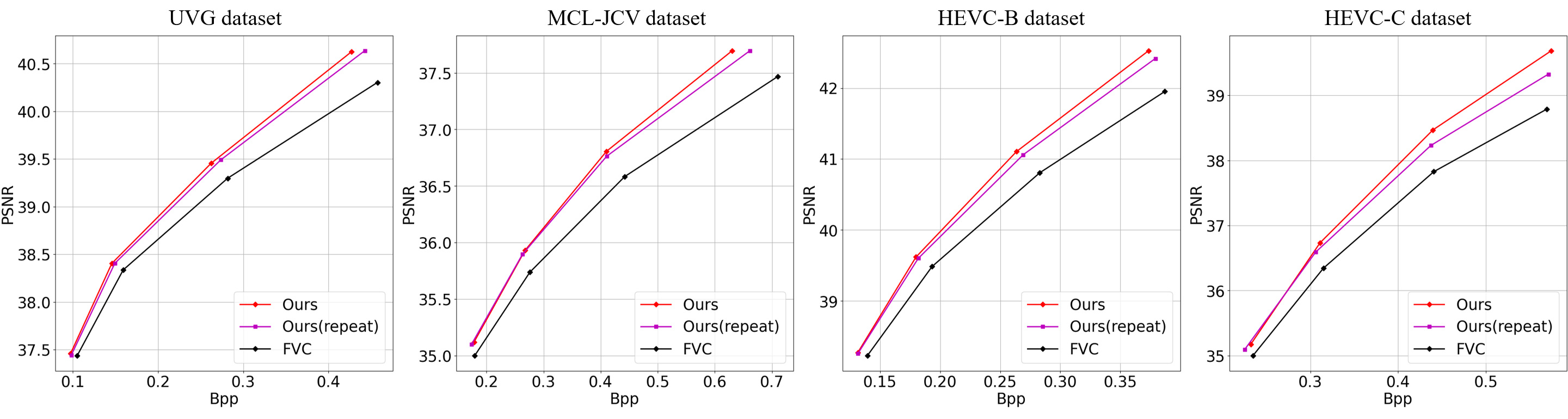}
   \end{center}  
   \caption{RD-curve comparison using FVC.}
   \label{fig:fvc}
\end{figure}

We applied our methods to another baseline, FVC ~\cite{hu2021fvc}, an end-to-end video codec.
FVC only compresses P-frames, using x265 for I-frame compression. As shown in ~\cref{fig:fvc}, our method outperforms the baseline in benchmark datasets. 
This demonstrates that our proposed approach performs well even in different architectures, and also can be applied to P-frame settings.

\subsubsection{Comparison to the previous methods}
We compared our approach to both previous methods applied to video compression, such as bias-tuning and encoder fine-tuning, as well as methods originally designed for other tasks, including domain transfer and image compression. \cref{tab:bdrate_reb} shows the BD-rate on the UVG dataset, comparing our method to x265 with bias-tuning, encoder fine-tuning, and two other methods: Tsubota \etal ~\cite{tsubota2023universal} (adapter using matrix decomposition) and Shen \etal ~\cite{shen2023dec} (adapter using depthwise separable convolution and activation functions). We adapted these other-task methods to the video compression context, and our method demonstrated the best performance among all PEFT methods. This underscores the effectiveness of our proposed architectural design.

\begin{table}[t]
\centering
\resizebox{\columnwidth}{!}{%
\setlength{\tabcolsep}{5pt}
\begin{tabular}{@{}ccccccc@{}}
\toprule
            & Ours    & Ours (repeat) & Bias & Encoder &  Tsubota \etal & Shen \etal \\ \midrule
BD-rate (\%)        & -6.48 & -0.14   & 65.84  & 13.24 & 9.47 & 11.69   \\

\bottomrule
\end{tabular}%
}

\caption{BD-rate comparison with previous methods using anchor as x265}
\label{tab:bdrate_reb}
\end{table}

\section{Limitations} 
\label{sec:Limitation}

The models currently do not represent state-of-the-art (SOTA) models. Specifically, the currently best-performing models such as DCVC series~\cite{li2023neural, li2024neural} do not provide the training code or details, and thus, it was not feasible to train them using conventional methods, thereby hindering applying our method to those SOTA models. 
We have shown the effectiveness of our methods on two widely acknowledged NVCs. We believe more investigation to adapt our method to the recent models would be beneficial in enhancing the performance and applicability.

\section{Conclusion}
\label{sec:Conclusion}

We introduce a novel parameter-efficient instance-adaptive method, which is adapted to scale-space flow models in both I-frame and P-frame settings. Our approach utilizes linear operations for reparameterization, ensuring no additional latency during decoding. Training factorized kernels with duplications maintain performance while reducing the number of bits transmitted in lower bpp areas, as evidenced by various BD-rate measurements. Training these kernels without duplications yields superior performance in RD-curves. Additional experimental results demonstrate the robustness, speed, and generalization capabilities of our methods. We believe that our work represents a significant step towards enhancing the efficiency and adaptability of instance-adaptive video compression.

\subsubsection{\ackname}
This research was supported in parts by the grant (RS-2023-00245342) from the Ministry of Science and ICT of Korea through the National Research Foundation (NRF) of Korea, Korea Institute of Energy Technology Evaluation and Planning (KETEP) and the Ministry of Trade, Industry \& Energy (MOTIE) of the Republic of Korea (No. 20224000000360), and Institute of Information \& communications Technology Planning \& Evaluation (IITP) grant funded by the Korea government (MSIT) (RS-2019-II190421, Artificial Intelligence Graduate School Program (Sungkyunkwan University)).

%
%
\bibliographystyle{splncs04}
\bibliography{main}
\clearpage
\newcommand{\beginsupplement}{
\setcounter{section}{0}
}

\renewcommand\thesection{\Alph{section}}
\begin{center}
\noindent \large\textbf{Supplementary Materials}\\

\end{center}
\beginsupplement

\section{Trainable parameters}
\label{sec:params}
As shown in \cref{tab:num_params}, our proposed methods use far fewer training parameters than the full fine-tuning method. This is mainly because our methods do not have parameters in the encoder and hyperprior parts that need training, allowing for a faster learning process. Additionally, our methods only need to update a relatively small set of parameters, increasing their overall efficiency. Specifically, the method that repeats parameters has a much smaller number of parameters compared to the other methods. This points to the high efficiency and practicality of our methods and strengthens their potential for effective use in various real-world situations.

\section{Adapter on hyperprior model}
\begin{figure}[t]
  \centering
   \includegraphics[width=\linewidth]{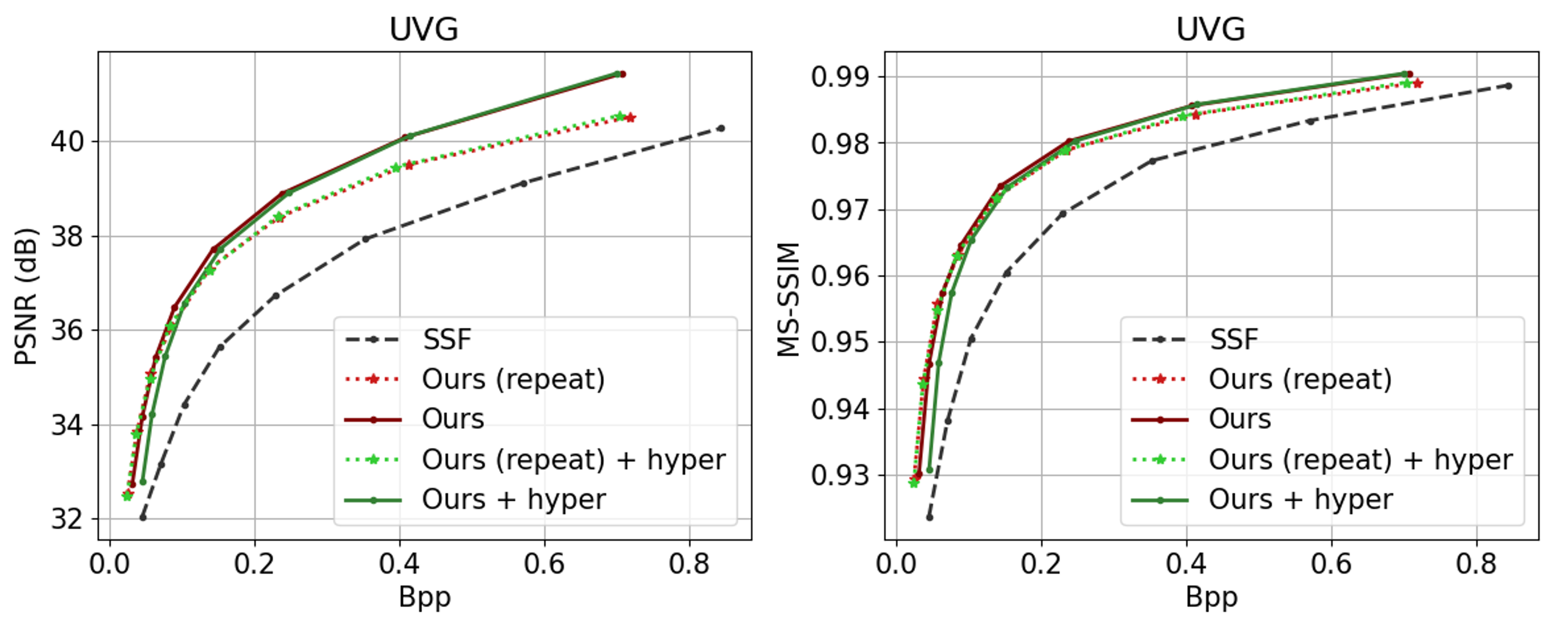}

   \caption{RD-curve when apply adapter on hyperprior model. Comparison conducted on UVG dataset.}
   \label{fig:hyper}
\end{figure}

We conducted experiments on integrating our adapter structure with hyperprior models. As \cref{fig:hyper} shows, this did not lead to improvements in PSNR and MS-SSIM values. Furthermore, the compression performance got worse at lower bitrates. This can be attributed to the fact that the added structure makes more bits need to be transferred.

\section{GoP size variation}
\begin{figure*}[ht!]
  \centering
   \includegraphics[width=\linewidth]{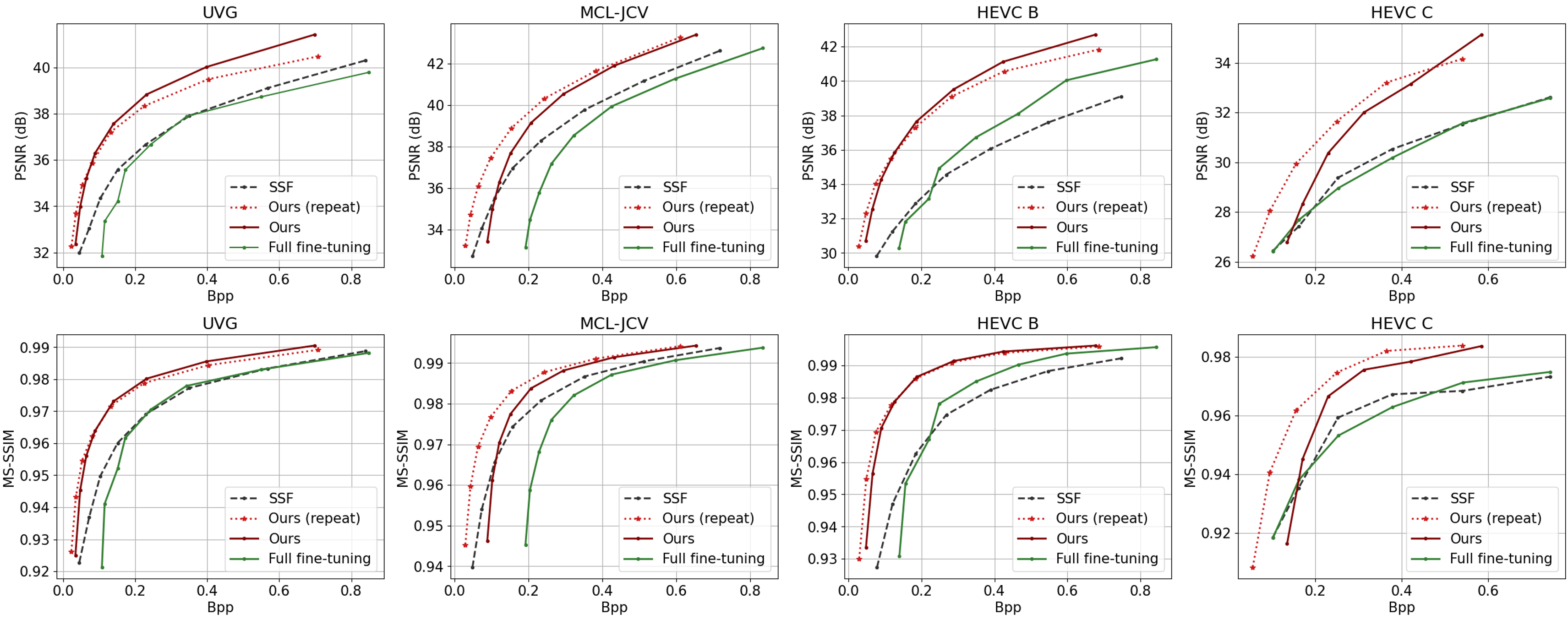} 
   \caption{RD-curve with GoP set to 32. Comparison conducted on UVG, MCL-JCV, HEVC class B, and C datasets.}
   \label{fig:rdcurve32}
\end{figure*}

As demonstrated in prior researches ~\cite{li2022hybrid,li2023neural}, the commonly used practical Group of Pictures (GoP) size is close to 32. Thus, We also evaluated the performance with the GoP size set to 32, using the same Rate-distortion (RD) curve. The datasets for this evaluation remained the same, including UVG ~\cite{mercat2020uvg}, MCL-JCV~\cite{wang2016mcl}, and HEVC class B and C ~\cite{bossen2013common}. As shown in \cref{fig:rdcurve32}, the RD curves of PSNR and MS-SSIM displayed largely similar performance to the previous result with a smaller GoP size. Although there was a slight decrease in performance at lower bitrates, the overall performance remained consistent, demonstrating the robustness and applicability of our proposed methods to larger GoP sizes. This suggests that our proposed methods can be effectively applied even when the GoP size is increased, further enhancing the versatility of our method.

\section{Qualitative results}
\begin{figure*}[ht!]
  \centering
   \includegraphics[width=0.8\linewidth]{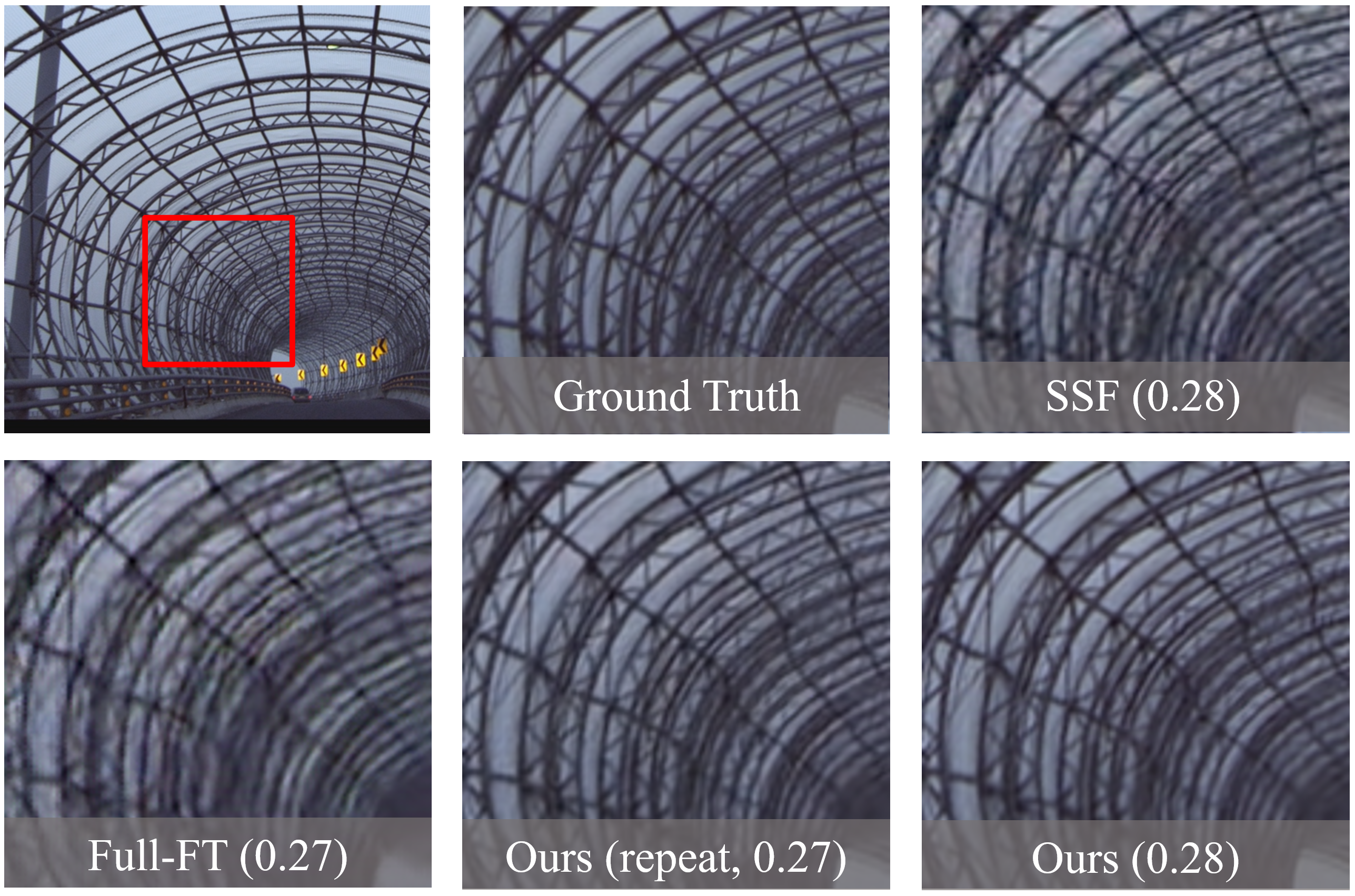} 
   \caption{Qualititive results of MCL-JCV 10 dataset.}
   \label{fig:qual_mcl10}
\end{figure*}

\begin{figure*}[ht!]
  \centering
   \includegraphics[width=0.8\linewidth]{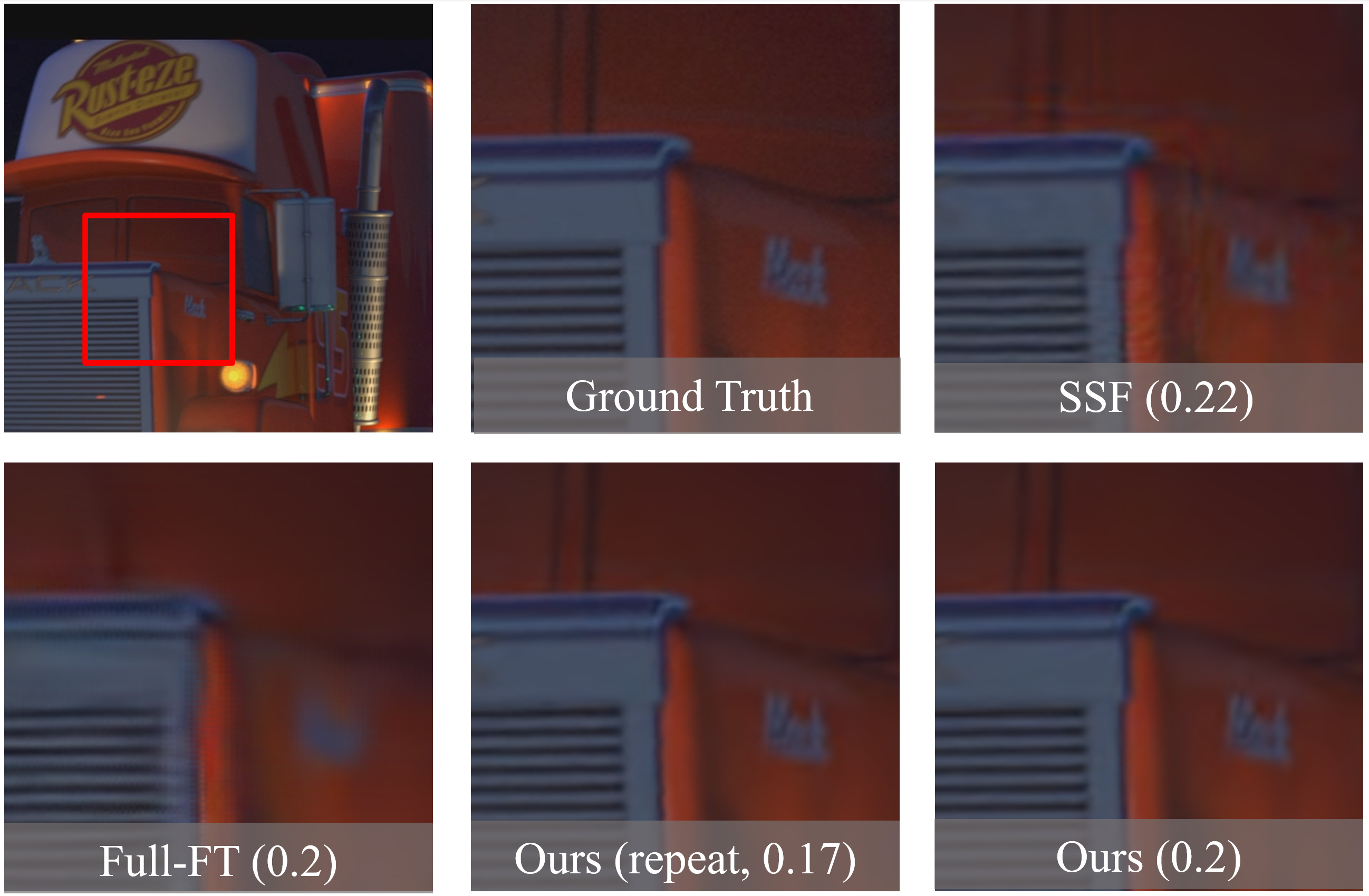} 
   \caption{Qualititive results of MCL-JCV 24 dataset.}
   \label{fig:qual_mcl24}
\end{figure*}

\begin{figure*}[ht!]
  \centering
   \includegraphics[width=0.8\linewidth]{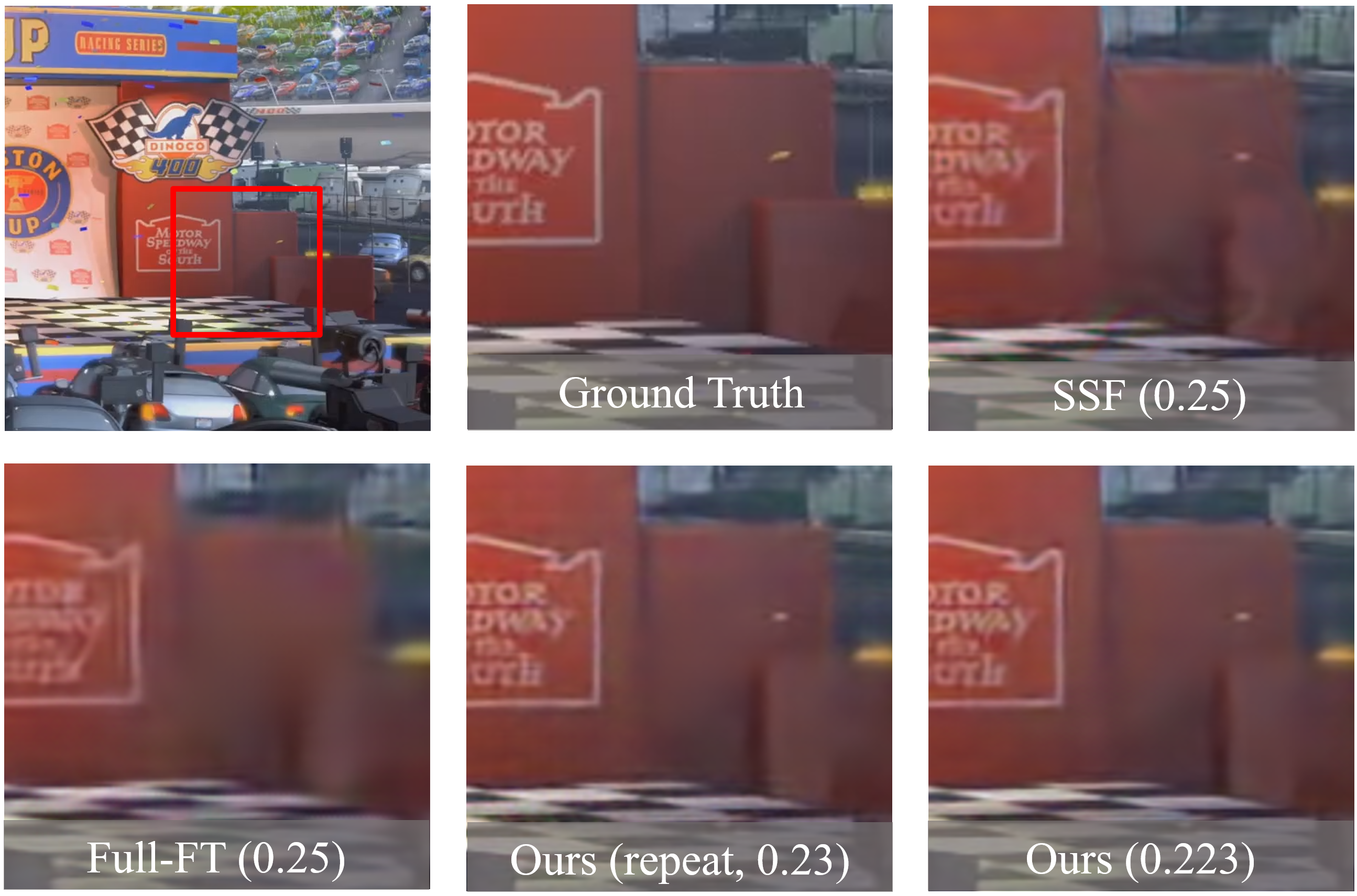} 
   \caption{Qualititive results of MCL-JCV 25 dataset.}
   \label{fig:qual_mcl25}
\end{figure*}

As indicated in \cref{sec:ablation study}, some datasets, especially those with cartoon-style or complex movements, pose challenges in reconstructing images. Therefore, we present the qualitative results for each dataset in \cref{fig:qual_mcl10}, \cref{fig:qual_mcl24}, and \cref{fig:qual_mcl25}. The comparison is made at similar bpp settings, revealing that SSF~\cite{agustsson2020scale} exhibits motion blur in complex domains. Moreover, full fine-tuning results in distortion from the original, failing to accurately represent finer details. In contrast, both of our methods can effectively represent their respective areas without motion blur, even in the cartoon domain. These qualitative results highlight the superior overfitting mitigation capability of our methods.

\section{PSNR per frame}
\begin{figure*}[ht!]
  \centering
  \begin{tabular}{c}
    \includegraphics[width=\linewidth]{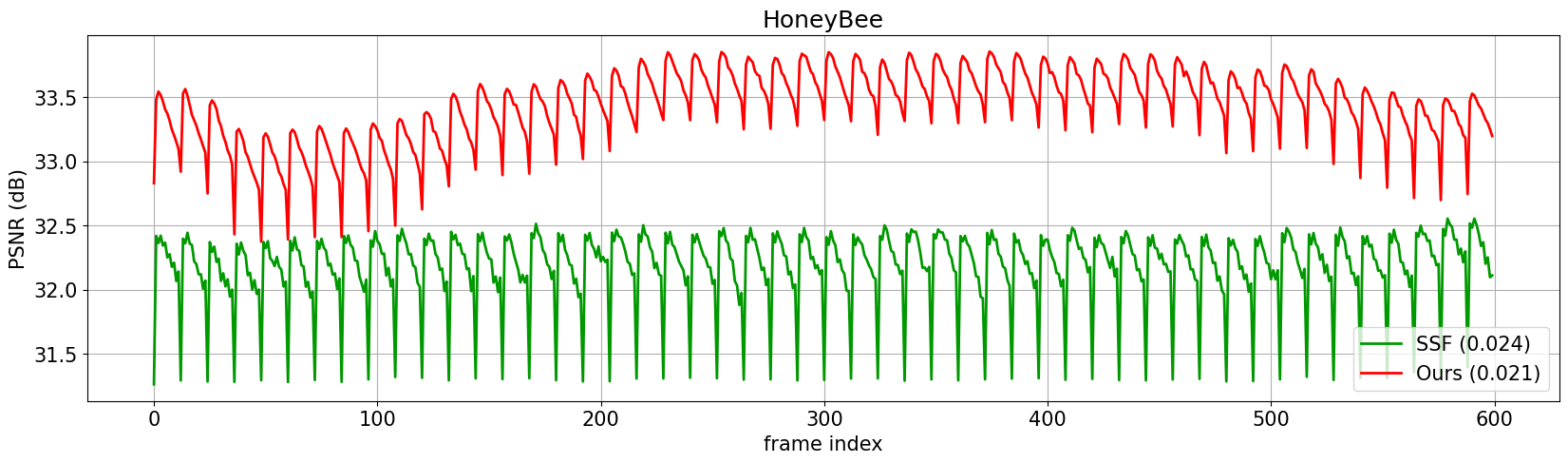} \\
    \includegraphics[width=\linewidth]{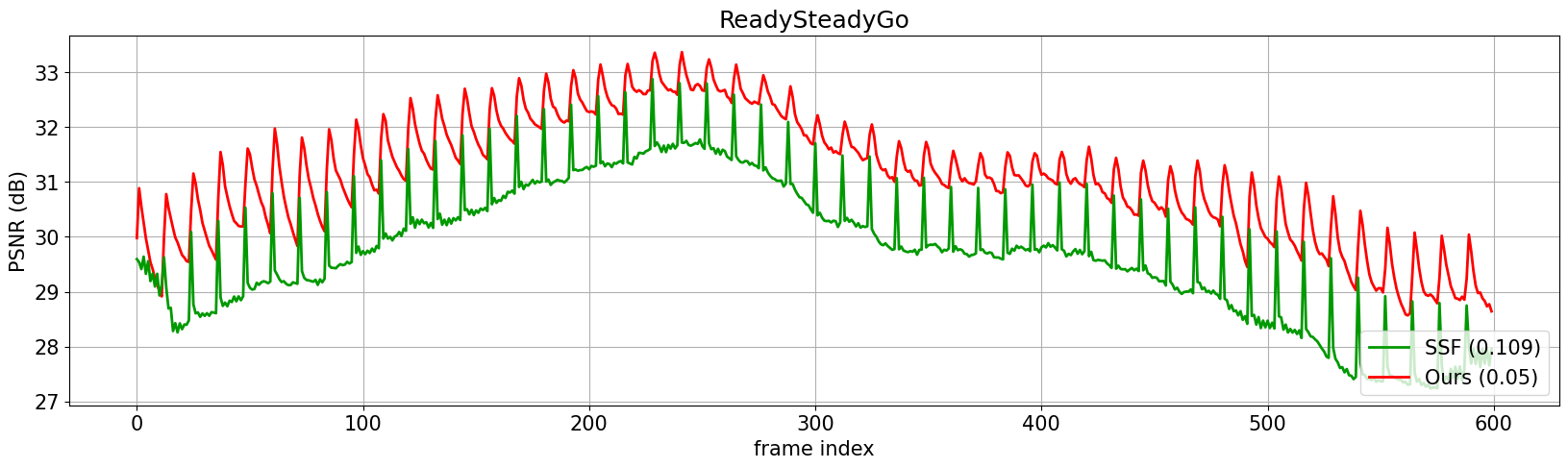}\\
    \includegraphics[width=\linewidth]{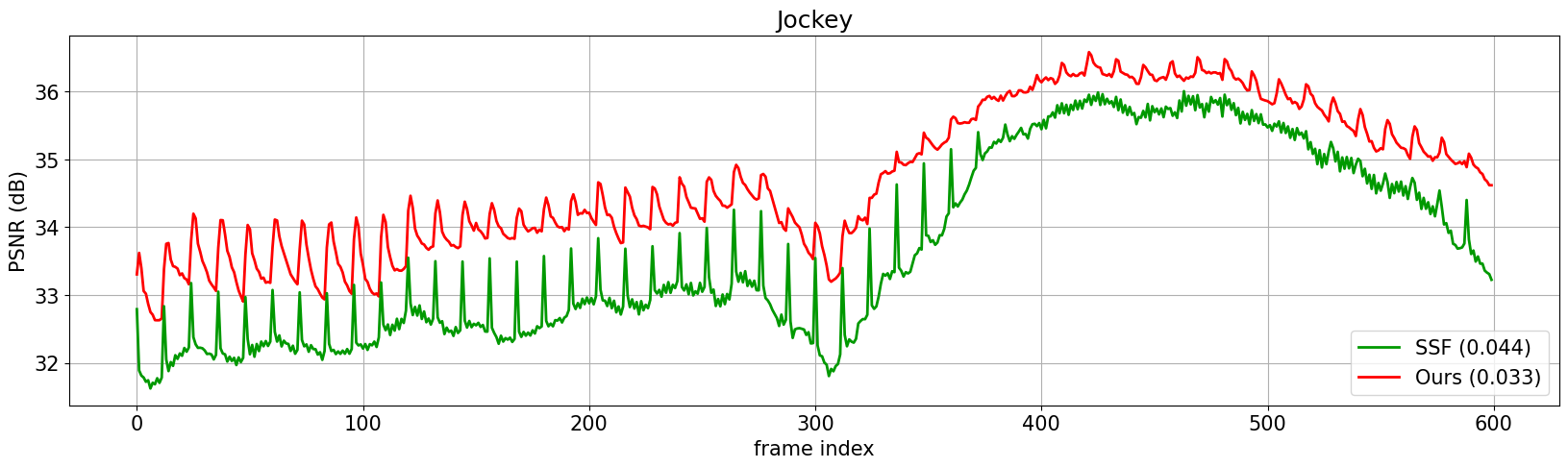} 
  \end{tabular}
   \caption{PSNR for each frame using the same baseline model, tested on the `HoneyBee', `ReadySteadyGo', and `Jockey' sequence. The Number in the legend represent bpp. }
   \label{fig:frame}
\end{figure*}

To assess the detailed performance of our method, we measure the PSNRs for each frame. As depicted in \cref{fig:frame}, the comparison is made between the baseline and our method with no duplication, focusing on the some of UVG dataset sequence. Both the baseline and our method show an increasing trend in PSNR. However, our method exhibits an overall improvement in PSNRs of approximately 1 dB, with smaller spikes, even though bpp is lower than baseline. This suggests that our method may be prone to overfitting the input video sequences with saving the number of bits.

\begin{table}[t]

\resizebox{\columnwidth}{!}{%
\setlength{\tabcolsep}{5pt}
\begin{tabular}{c|c|cccc}
\toprule
\multirow{2}{*}{} & \multirow{2}{*}{Total params. (M)} & \multicolumn{4}{c}{Train params. (M)}                                                                      \\ \cline{3-6} 
                  &                        & \multicolumn{1}{c|}{Encoder}  & \multicolumn{1}{c|}{Hyperprior} & \multicolumn{1}{c|}{Decoder} & all      \\
\midrule
Full fine-tuning  & 34.24               & \multicolumn{1}{c|}{12.70} & \multicolumn{1}{c|}{16.59}   & \multicolumn{1}{c|}{4.95} & 34.24 \\
Ours              & 35.03               & \multicolumn{1}{c|}{0}        & \multicolumn{1}{c|}{0}          & \multicolumn{1}{c|}{0.79}  & 0.79   \\
Ours(repeat)      & 34.27               & \multicolumn{1}{c|}{0}        & \multicolumn{1}{c|}{0}          & \multicolumn{1}{c|}{0.03}   & 0.03    \\

\bottomrule

\end{tabular}%
}
\caption{Number of training parameters for video sequence instance-adaptation. }
\label{tab:num_params}
\end{table}

\end{document}